\documentclass[prd,aps,nofootinbib,notitlepage,showpacs,showkeys,preprintnumbers]{revtex4-1}
\usepackage{graphicx,epsf,amsmath,amsfonts,amssymb,amsbsy}
\usepackage{epsfig}
\usepackage{epstopdf}

\newcommand{\ds}{\displaystyle}
\newcommand{\vev}[1]{\langle#1\rangle}
\newcommand{\mat}{\left ( \begin{array}}
\newcommand{\emat}{\end{array} \right )}
\newcommand{\vect}{\left ( \begin{array}{c}}
\newcommand{\evect}{\end{array} \right )}

\begin{document}
\title{
Dense baryonic matter with chiral imbalance and charged pion condensation \footnote{Presented at Excited QCD 2020}
}
\author{T. G. Khunjua $^{1),~2)}$, K. G. Klimenko $^{3)}$, and R. N. Zhokhov $^{3),~4)}$ }

\affiliation{$^{1)}$ The University of Georgia, GE-0171 Tbilisi, Georgia}
\affiliation{$^{2)}$ Department of Theoretical Physics, A. Razmadze Mathematical Institute, I. Javakhishvili Tbilisi State University, GE-0177 Tbilisi, Georgia}
\affiliation{$^{3)}$ State Research Center
of Russian Federation -- Institute for High Energy Physics,
NRC "Kurchatov Institute", 142281 Protvino, Moscow Region, Russia}
\affiliation{$^{4)}$  Pushkov Institute of Terrestrial Magnetism, Ionosphere and Radiowave Propagation (IZMIRAN),
108840 Troitsk, Moscow, Russia}

\begin{abstract}
After the first predictions in seventies that in dense nuclear matter, for example, in neutron stars there could be pion condensation phenomenon, it was shown that at least s-wave pion condensation is unlikely to occur in such a medium in different approaches including NJL model consideration with neutrality and $\beta$-equilibrium condition. Then it has been found a condition that can promote this phenomenon in dense baryonic matter, it is chiral imbalance, so now it is interesting if this generation can survive rather strict requirement of electric neutrality and $\beta$-equilibrium. Here in this paper the generation of pion condensation in dense baryonic matter by chiral imbalance is investigated. It is shown that electric neutrality and $\beta$-equilibrium conditions do not spoil this phenomenon and this leads to interesting applications for physics of neutron stars. There are several possible mechanisms of generation of chiral imbalance in dense matter. In a view of latest and forthcoming NICER results and the first observed and possibly new events of neutron star mergers it is rather interesting to explore how possible chiral imbalance in neutron star can influence and change EOS and, hence, mass-radius relation. It is a first step in that direction.
\end{abstract}
\maketitle
  
\section{Introduction}

It is anticipated that the phase structure of QCD at finite temperature and baryon density is rather rich. The hot and dense QCD studies has not just pure academic interest but is relevant for various physical situation, for example, early universe, physics of neutron star and its mergers, heavy-ion collisions etc.
The conjecture that at high densities of nuclear matter there could be such a phenomenon as pion condensation (PC) was considered a long time ago in \cite{MigdalKogutpion}.  Phase structure at isospin imbalance has been studied in lattice QCD approach or different QCD-like effective models in Refs \cite{Son, koguthe,
abuki,ak}). It was shown in these papers that if there is an isospin imbalance then charged PC phenomenon can be generated in quark matter. It was shown in \cite{eklim} that there is PC in dense quark matter but later it was displayed that if the electric charge neutrality constraint is imposed, the charged PC phenomenon depends strongly on the bare (current) quark mass and forbidden for physical quark masses \cite{abuki}. However, several external factors, including chiral imbalance \cite{ekk},  promoting the formation of the charged PC phase in dense quark matter have recently been discovered \cite{ekk, ekkz}. But in these papers the influence of chiral imbalance on the formation of the charged PC phenomenon  was considered without taking into account the possible electrical neutrality and $\beta$-equilibrium of the medium, 
i.e. the results of these studies are not directly applicable to such astrophysical objects as neutron stars, etc. In the current letter we tried to fill this gap and study charged PC phenomenon in electric neutral dense quark matter in $\beta$-equilibrium. 

\section{The model}

Let us consider matter of the compact star, i. e. the system composed of $u$ and $d$ quarks and electrons and described by the following Lagrangian:
\begin{eqnarray}
&& {\cal L}=\bar q\Big [\gamma^\nu\mathrm{i}\partial_\nu-m_0
+\bar\mu\gamma^0+\nu\tau_3\gamma^0
+\nu_{5} \tau_3\gamma^0\gamma^5+\mu_{5}\gamma^0\gamma^5\Big ]q
\nonumber\\&&
+G\Big [(\bar qq)^2+(\bar q\mathrm{i}\gamma^5\vec\tau q)^2 \Big
]+\bar\psi\Big(\gamma^\nu\mathrm{i}\partial_\nu-\mu_{Q}\gamma^0 \Big )\psi,\label{60}~~~60
\end{eqnarray}
where where $q(x)$ is $q=(q_u,q_d)^T$ and $q_u$ and $q_d$ are four-component Dirac spinors corresponding to $u$ and $d$ quarks and $\psi(x)$ is a Dirac spinor of electrons,  $\tau_k$ ($k=1,2,3$) are Pauli matrices. For simplicity, we assume that the mass of electron is equal to zero. $\mu_B$, $\mu_Q$, $\mu_{I5}$ and $\mu_{5}$ (for simplicity the notations $\bar\mu\equiv\mu+\frac{\mu_Q}6$,~~$\mu=\frac{\mu_B}{3}$,~~ $\nu\equiv\frac{\mu_Q}2,~~\nu_5\equiv\frac{\mu_{I5}}2$ are used) are baryon, electric charge, chiral isospin and chiral chemical potentials, they are connected with particle number chemical potentials of $u$ and $d$ quarks and electrons, $\mu_u,\mu_d$ and $\mu_e$ in the following way $\mu_u=\frac{\mu_B}{3}+\frac{2\mu_Q}{3},~\mu_d=\frac{\mu_B}{3}-\frac{\mu_Q}{3},~\mu_e=-\mu_Q$

 These chemical potentials are conjugate to baryon, electric charge, chiral isospin and chiral densities,
$n_B\equiv\frac 13\bar q\gamma^0q=\frac 13(n_u+n_d),~~ n_I\equiv\frac 12\bar q\gamma^0\tau^3 q=\frac 12(n_u-n_d), n_{Q}\equiv\bar q\widetilde Q\gamma^0q-|e|\bar\psi\gamma^0\psi=\frac{2|e|}{3}n_u-\frac{|e|}{3}n_d-|e|n_e$,
(where $n_u=\bar q_u\gamma^0q_u$, $n_d=\bar q_d\gamma^0q_d$ and $n_e=\bar\psi\gamma^0\psi$ are particle number densities of $u$ and $d$ quarks and electrons). 

According to recent studies \cite{andrianov}, chiral imbalance can appear in dense quark medium. We will study dense cold quark matter under two constraints, the first one is $\beta$ equilibrium meaning all $\beta$ processes such as $d\rightarrow u+e+\bar\nu_{e}$, $u+e\rightarrow d+\nu_{e}$ go with equal rates in both directions and leading to the relation on chemical potentials $\mu_d=\mu_u+\mu_e-\mu_{\nu}$ (as neutrinos leave the system, we assume that  $\mu_{\nu}=0$).
The second one is requirement of local electrical neutrality, $\vev{n_Q}=0$, that constrain the value of electric charge chemical potential $\mu_{Q}$. 
%
%

It is more convenient to use a semibosonized version of the Lagrangian, which contains
composite bosonic fields $\sigma (x)$ and $\pi_a (x)$ $(a=1,2,3)$:
\begin{eqnarray}
\widetilde {\cal L}\ds &=&\bar qDq 
-\frac{1}{4G}\left (\sigma\sigma+\pi_a\pi_a\right )
+ \bar\psi\Big (\gamma^\nu\mathrm{i}\partial_\nu -\mu_{Q}\gamma^0 \Big )\psi,\label{2}
\end{eqnarray}
where $D\equiv\gamma^\rho\mathrm{i}\partial_\rho
+\mu\gamma^0
+ \nu\tau_3\gamma^0+\nu_{5}\tau_3\gamma^0\gamma^5+\mu_5\gamma^0\gamma^5-m_{0}-\sigma
-\mathrm{i}\gamma^5\pi_a\tau_a$.
In the following we assume that $N_c=3$.

From the auxiliary Lagrangian (\ref{2}) one gets the equations
for the bosonic fields
$\sigma(x)=-2 G(\bar qq);~\pi_a (x)=-2 G(\bar q
\mathrm{i}\gamma^5\tau_a q)$.
The bosonic field $\pi_3 (x)$ can be identified with neutral pion $\pi^0(x)$, whereas  $\pi_{1, 2} (x)$ with the charged pion $\pi^\pm (x)$ in the following combinations $\pi^\pm (x)=(\pi_1 (x)\mp i\pi_2 (x))/\sqrt{2}$.   
If $\vev{\pi_1(x)}\ne 0$ and/or $\vev{\pi_2(x)}\ne 0$ and one has non-zero condensates $\pi^+(x)$ and $\pi^-(x)$, this phase is usually called the charged pion condensation (PC) phase. In addition, these condensates are not invariant with respect to the electromagnetic $U_Q(1)$ transformations and 
this phase is superconducting.
Futhermore, we suppose that the ground state expectation values $\vev{\sigma(x)}$ and $\vev{\pi_a(x)}$ do not depend on spacetime coordinates $x$,
$\vev{\sigma(x)}\equiv M-m_{0},~\vev{\pi_1(x)}= \Delta,~\vev{\pi_{2,3}(x)}= 0$

One can obtain in the 
mean field approximation the expression for the thermodynamic potential (TDP) $\Omega (M,\pi_a)$, baryon charge $\vev{n_{B}}$ and electric charge $\vev{n_Q}$ densities.

\section{Phase diagram}
First let us recapitulate the story with charged pion condensation in dense quark matter (PC$_{d}$ phase).
The charged pion condensation in dense quark matter with isospin imbalance was predicted in the framework of NJL model in the chiral limit \cite{eklim}. It has been shown that at non-zero baryon density one can have PC in the system. 
Then it was demonstrated 
that charged PC phase is forbidden in electrically neutral dense quark matter with
$\beta$-equilibrium when current quark masses are close to their physical value \cite{abuki}.
After that there have been found several realistic conditions that create or enhance the generation of PC phase in the system. 
One can find the detailed discussion in \cite{Khunjua:2019nnv}.
 In particular it was shown that chiral imbalance can lead to the generation of PC$_{d}$ phase.

\begin{figure}
\includegraphics[width=0.49\textwidth]{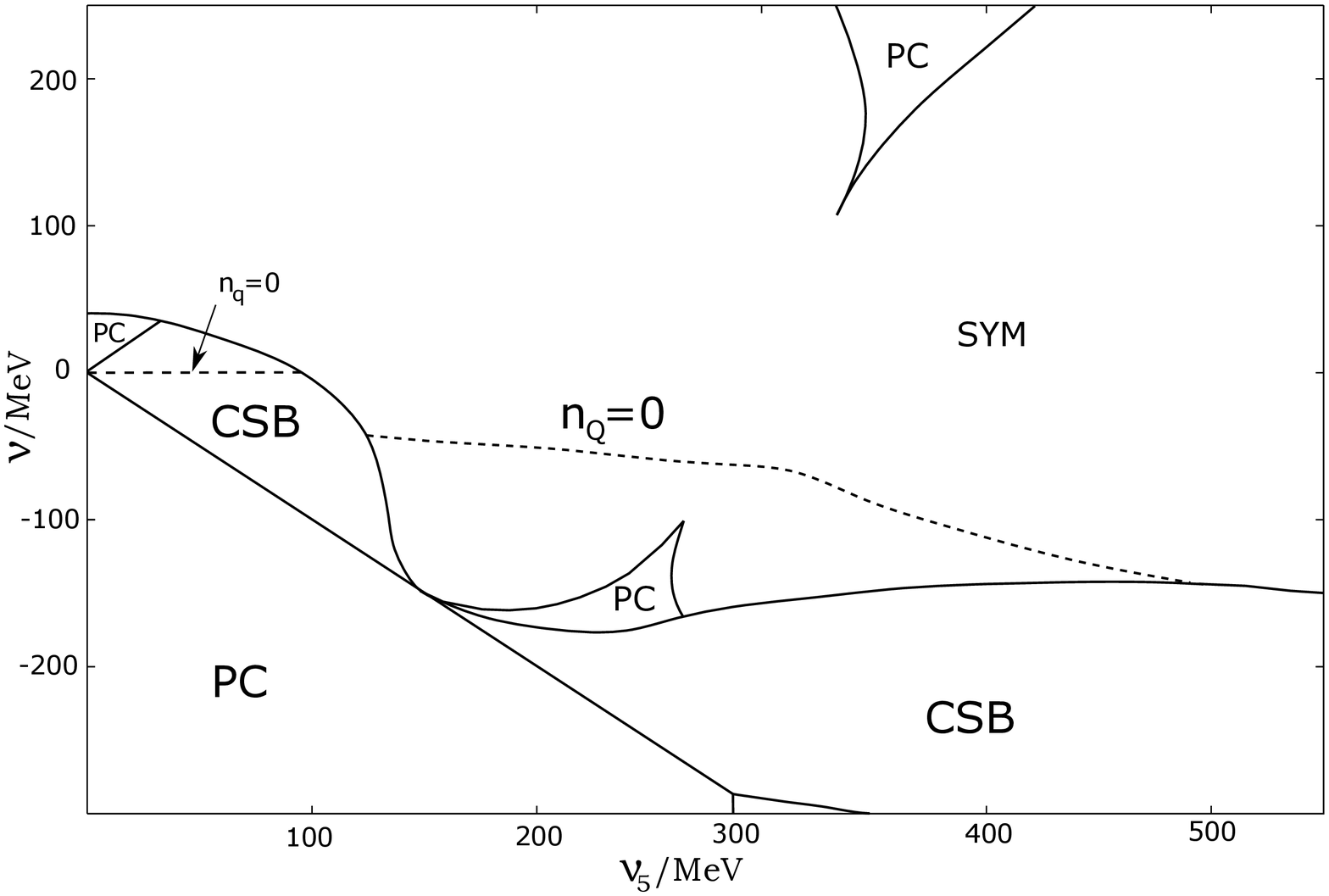}
 \hfill
\includegraphics[width=0.49\textwidth]{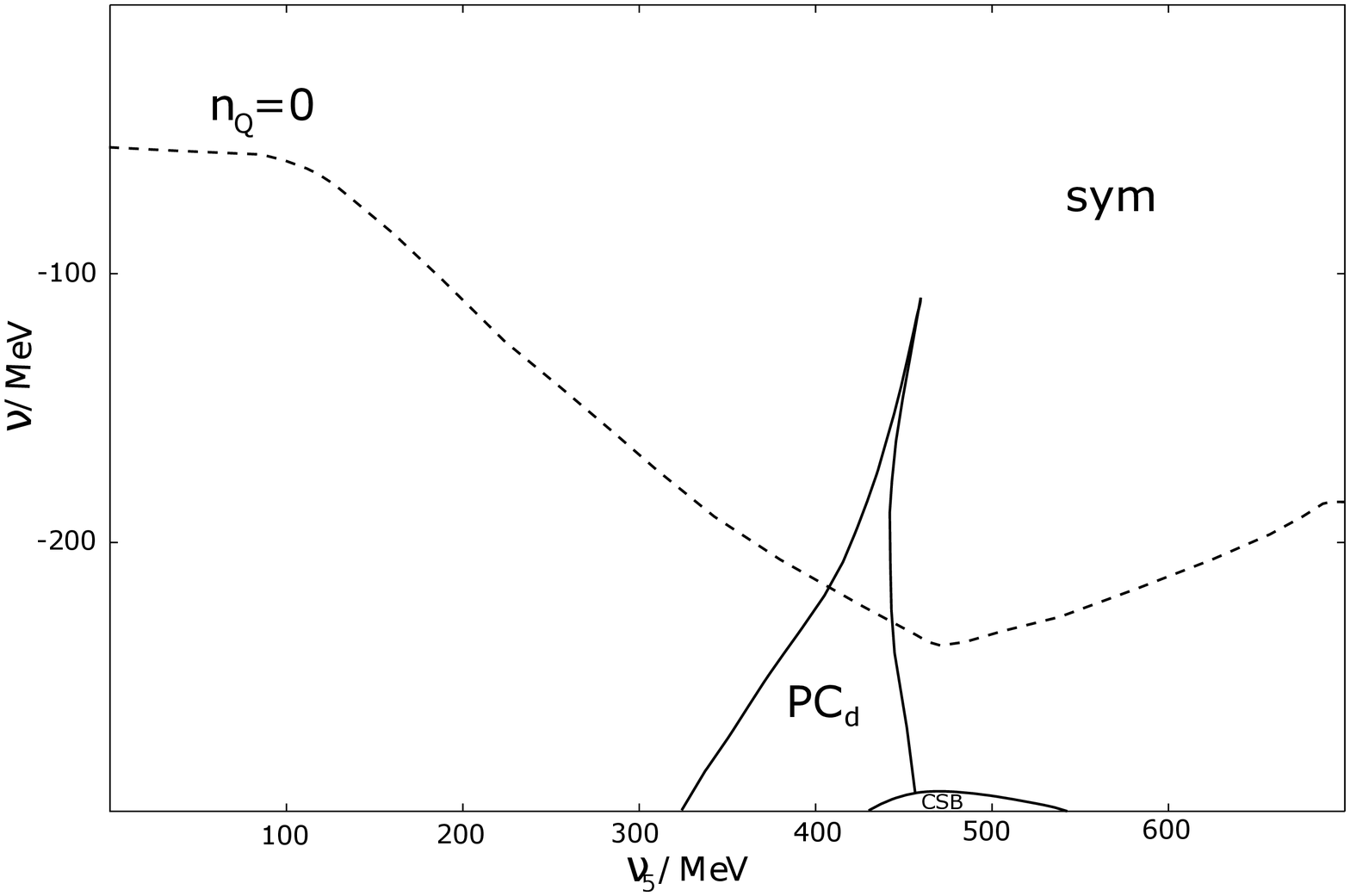}\\
\label{fig7}
\parbox[t]{0.45\textwidth}{
 \caption{ $(\nu_{5},\nu)$ phase diagram at $\mu=0$ MeV and $\mu_{5}=0$ MeV. 
 }
 }\hfill
\parbox[t]{0.45\textwidth}{
\caption{  $(\nu_{5},\nu)$ phase diagram at $\mu=500$ MeV and $\mu_{5}=0$ MeV. 
} }
\label{fig8}
\end{figure}

Since in dense baryonic (quark) matter with chiral imbalance the charged PC can be easily generated it is interesting to check if this effect can also be realized in neutral dense baryonic (quark) matter with $\beta$-equilibrium or the neutrality condition would completely destroy this effect.

First let us start with the particular case of zero chiral imbalance $\mu_{5}=0$ and only $\nu_5\neq0$. One can see at Fig. 1 that there is no  PC$_{d}$ at $\mu=0$. 
The charge neutrality line starts to cross the PC$_{d}$ phase from value of quark chemical potential $\mu=380$ MeV at values of chiral isospin chemical potential of several hundred MeV. First the crossed region of the PC$_{d}$ phase is rather small it but increases with increasing quark chemical potential and one can see at Fig. 2 that at the value of $\mu=500$ MeV the PC$_{d}$ phase is realized at values of chiral isospin chemical potential $\nu_{5}$ from 400 MeV up to 450 MeV 
(rather large region of order of 50 MeV). These values are still less than the cut-off and in the scope of validity of the model. One can already see that the generation of charged pion condensation in dense quark matter by chiral imbalance predicted in \cite{ekk, ekkz} is realized in neutral matter with $\beta$-equilibrium. But now let us turn to the case when there is non-zero chiral imbalance in two forms, chiral $\mu_{5}$ and chiral isospin $\nu_{5}$ chemical potentials are taken into account.  In this case one can see that even at rather moderate values of $\mu_{5}$ the generation of PC$_{d}$ phase is greatly enhanced, for example, one can see in Fig. 3  that at $\mu_{5}=150$ MeV 
in the system there realizes the PC$_{d}$ phase for range of $\nu_{5}$ from 200 to 370 MeV.
And if there is a  rather large chiral imbalance  $\mu_{5}$ in the system the  PC$_{d}$ phase appears for a very large parameter space, in Fig. 4 the phase diagram is depicted at $\mu_{5}=300$ MeV and one can see that for a rather large range of values (270 MeV) of $\nu_{5}$ there realizes the PC$_{d}$ phase in the system.

\begin{figure}
\includegraphics[width=0.49\textwidth]{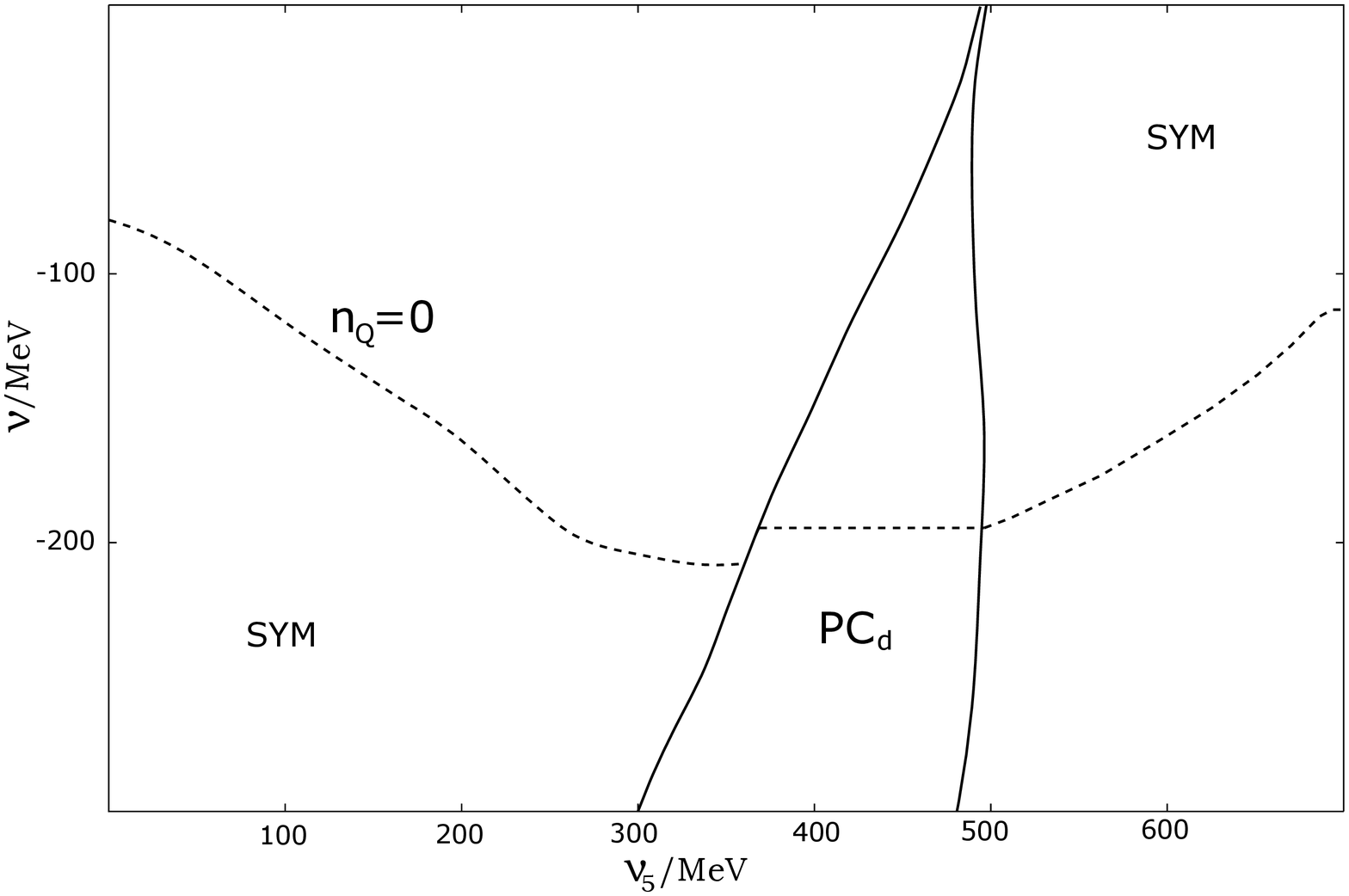}
 \hfill
\includegraphics[width=0.49\textwidth]{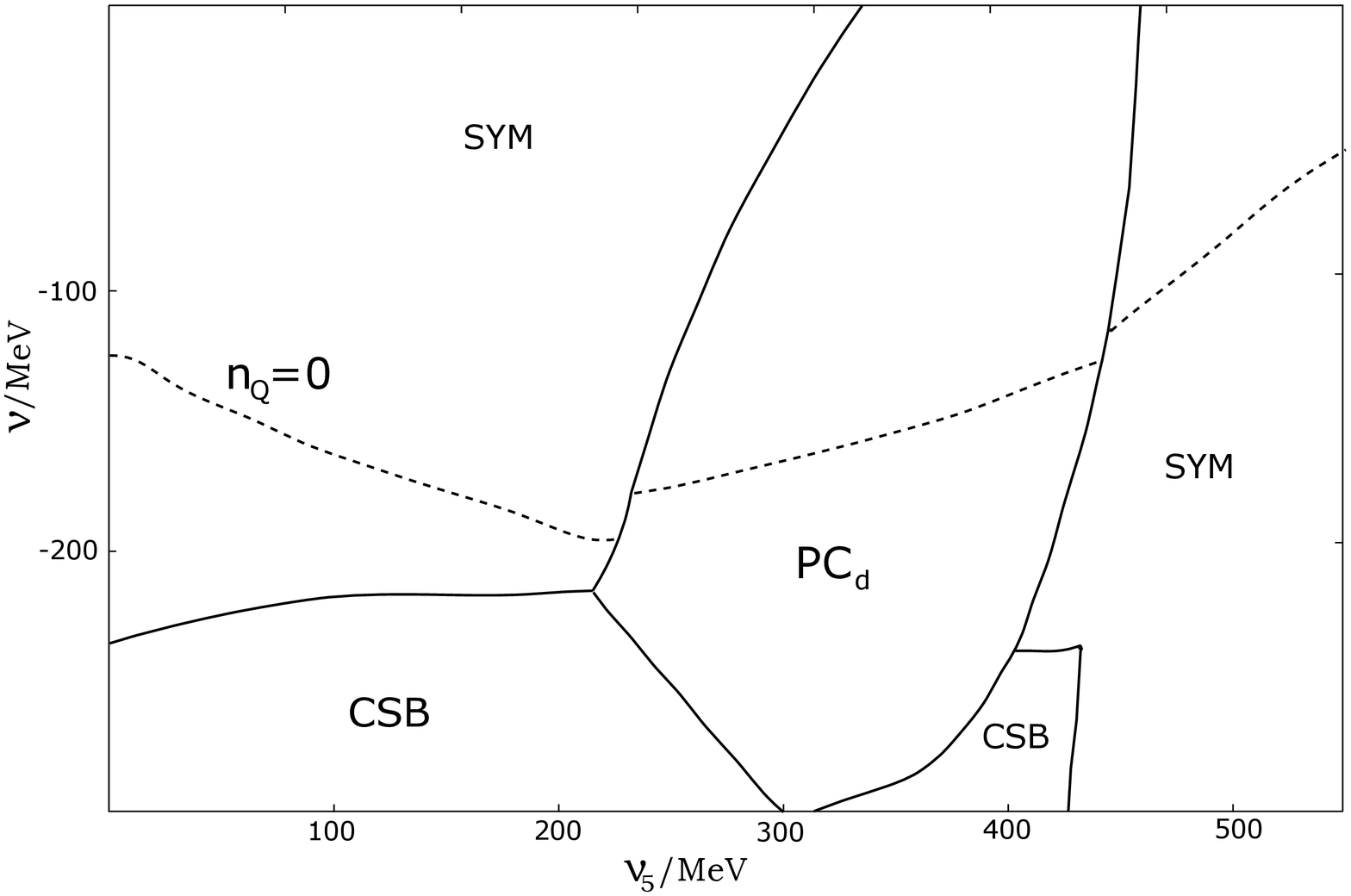}\\
\label{fig14}
\parbox[t]{0.45\textwidth}{
 \caption{  $(\nu_{5},\nu)$ phase diagram at $\mu=500$ MeV and $\mu_{5}=150$ MeV.  
 }
 }\hfill
\parbox[t]{0.45\textwidth}{
\caption{  $(\nu_{5},\nu)$ phase diagram at $\mu=400$ MeV and $\mu_{5}=300$ MeV. 
} }
\label{fig15}
\end{figure}
\section{Summary and Conclusions}
In this paper the question if the generation of pion condensation in dense baryonic (quark) matter by chiral imbalance remains valid in the case of matter in neutron stars, i. e. if one accounts for charge neutrality and $\beta$-equilibrium conditions. It has been demonstrated that chiral imbalance even in one form, chiral isospin chemical potential, generates significant region of PC$_{d}$ phase for almost any significantly large value of baryon density ($\mu>380$ MeV).
Furthermore, it has been shown that if there is chiral imbalance in both forms, chiral isospin and chiral chemical potentials, then the PC$_{d}$ phase is very easily generated in the system. Hence, one can conclude that chiral imbalance very effectively generates pion condensation in dense matter in neutron star conditions.

{\bf Acknowledgements:} RZ is grateful for support of Russian Science Foundation under grant No 19-72-00077. The work is supported by Foundation for the Advancement of Theoretical Physics and Mathematics
BASIS.

\end{document}